\documentclass[conference]{IEEEtran}
\usepackage{amssymb,stmaryrd,amsmath,amsfonts,rotating}
\usepackage{color}
\usepackage{bm}

\newcommand{\dL}{L}
\newcommand{\dl}{d_{\mathrm{l}}}
\newcommand{\dk}{k}
\newcommand{\dr}{d_{\mathrm{r}}}

\allowdisplaybreaks
\begin{document} 
\title{Efficient Termination of Spatially-Coupled Codes}
\author{
Koji Tazoe, Kenta Kasai and Kohichi Sakaniwa\\
\authorblockA{
Dept. of Communications and Integrated Systems, 
Tokyo Institute of Technology, 152-8550 Tokyo, Japan.\\
Email: {\tt \{tazoe,kenta,sakaniwa\}@comm.ss.titech.ac.jp}} 
}

\maketitle
\begin{abstract}
Spatially-coupled  low-density parity-check  codes attract much attention due to 
their capacity-achieving performance and a memory-efficient sliding-window decoding algorithm.
On the other hand, the encoder needs to solve large linear equations to terminate the encoding process. 
In this paper, we propose modified spatially-coupled codes.
The modified $(\dl,\dr,L)$ codes have less rate-loss, i.e., higher coding rate, and have the same threshold as $(\dl,\dr,L)$ codes 
and are efficiently terminable by using an accumulator. 
\end{abstract}
\section{Introduction}
Spatially-coupled (SC) low-density parity-check (LDPC) codes attract much attention due to their capacity-achieving performance and a memory-efficient sliding-window decoding algorithm.
The studies on SC-LDPC codes date back to the invention of convolutional LDPC codes by Felstr{\"o}m and Zigangirov \cite{zigangirov99}. 
They introduced a construction method of $(\dl, \dr)$-regular convolutional LDPC codes from $(\dl, \dr)$-regular block LDPC codes. 
The convolutional LDPC codes exhibited better decoding performance than the underlying block LDPC codes under a fair comparison with respect to the code length. 
Lentmaier {\it et al.}~ observed that (4,8)-regular convolutional LDPC codes exhibited the decoding performance surpassing the belief propagation (BP) threshold of (4,8)-regular block LDPC codes \cite{lentmaier_II}. 
Further, the BP threshold coincides with the maximum a posterior (MAP) threshold of the underlying block LDPC codes with a lot of accuracy. 
Constructing convolutional LDPC codes from a block LDPC code improves the BP threshold up to the MAP threshold of the underlying codes. 

Kudekar {\it et al.}~ named this phenomenon  ``threshold saturation'' and proved rigorously for the binary-input erasure channel (BEC) \cite{5695130} and the binary-input memoryless output-symmetric (BMS) channels. \cite{2012arXiv1201.2999K}. 
In the limit of large $\dl,\dr,L$ and $w$, the SC-LDPC code ensemble $(\dl,\dr,L,w)$ \cite{5695130} was shown to {\em universally} achieve the Shannon limit of BMS channels under BP decoding. 
This means the transmitter does not need detail statistics of the channel but needs to know only the channel capacity. 
Such universality is not supported by other efficiently-decodable capacity-achieving codes, e.g., polar codes 
and irregular LDPC codes. 
According to the channel, polar codes need  frozen bit selection 
and irregular LDPC codes need optimization of degree distributions. 
We note that recently Aref and Urbanke proposed SC rateless codes \cite{ITW_AREF_URBANKE} which  are conjectured to universally achieve the capacity of BMS channels without knowing even the capacity of the channel at the transmitter. 

SC-LDPC codes are constructed from $L$ LDPC block codes and have lower coding rate than the underlying block codes. 
As increasing coupling number $L$, the rate-loss could be decreased and the block error rate are degraded. 
Kudekar et al. proposed to mitigate the rate-loss in \cite{DBLP:journals/corr/abs-1004-3742}. 

Pusane et al.  proposed an efficient encoding and decoding method of convolutional LDPC codes in \cite{pusane08TCOM}. 
Encoding of convolutional LDPC codes can be divided into two stages: sequential encoding and termination. 
The sequential encoding process can be efficiently calculated with computational cost $O(LM)$, where $M$ is the lifting number for protograph codes. 
On the other hand, termination process needs a multiplication of dense matrix of size, which needs computational cost $O(M^2)$. 
Since $M$ and $L$ are chosen so that $L\ll M$, termination needs more computations than decoding and sequential encoding. 

In this paper, we propose modified spatially-coupled codes.
The modified $(\dl,\dr,L)$ codes have less rate-loss, i.e., higher coding rate, and have the same threshold as $(\dl,\dr,L)$ codes 
and are efficiently terminable by using an accumulator. 
\begin{figure*}[t]
 \scriptsize{
 \begin{align*}
 \renewcommand\arraystretch{0.45}
&\left[ \begin{array}{@{\hspace{0mm}}l@{\hspace{0mm}}l@{\hspace{0mm}}l@{\hspace{0mm}}l@{\hspace{0mm}}l@{\hspace{0mm}}l@{\hspace{0mm}}l@{\hspace{0mm}}l@{\hspace{0mm}}l@{\hspace{0mm}}l@{\hspace{0mm}}l@{\hspace{0mm}}l@{\hspace{0mm}}l@{\hspace{0mm}}l@{\hspace{0mm}}l@{\hspace{0mm}}l@{\hspace{0mm}}l@{\hspace{0mm}}l@{\hspace{0mm}}l@{\hspace{0mm}}l@{\hspace{0mm}}l@{\hspace{0mm}}l@{\hspace{0mm}}l@{\hspace{0mm}}l@{\hspace{0mm}}l@{\hspace{0mm}}l@{\hspace{0mm}}l@{\hspace{0mm}}l@{\hspace{0mm}}l@{\hspace{0mm}}l@{\hspace{0mm}}l@{\hspace{0mm}}l@{\hspace{0mm}}l@{\hspace{0mm}}l@{\hspace{0mm}}l@{\hspace{0mm}}l}
P_{1,10}&P_{1,11}&P_{1,12}& & & & & & & & & & & & & & & & & & & & & & & & \\
P_{2,7}&P_{2,8}&P_{2,9}&P_{2,10}&P_{2,11}&P_{2,12}& & & & & & & & & & & & & & & & & & & & & \\
P_{3,4}&P_{3,5}&P_{3,6}&P_{3,7}&P_{3,8}&P_{3,9}&P_{3,10}&P_{3,11}&P_{3,12}& & & & & & & & & & & & & & & & & & \\
P_{4,1}&P_{4,2}&P_{4,3}&P_{4,4}&P_{4,5}&P_{4,6}&P_{4,7}&P_{4,8}&P_{4,9}&P_{4,10}&P_{4,11}&P_{4,12}& & & & & & & & & & & & & & & \\
 & & &P_{5,1}&P_{5,2}&P_{5,3}&P_{5,4}&P_{5,5}&P_{5,6}&P_{5,7}&P_{5,8}&P_{5,9}&P_{5,10}&P_{5,11}&P_{5,12}& & & & & & & & & & & & \\
 & & & & & &P_{6,1}&P_{6,2}&P_{6,3}&P_{6,4}&P_{6,5}&P_{6,6}&P_{6,7}&P_{6,8}&P_{6,9}&P_{6,10}&P_{6,11}&P_{6,12}& & & & & & & & & \\
 & & & & & & & & &P_{7,1}&P_{7,2}&P_{7,3}&P_{7,4}&P_{7,5}&P_{7,6}&P_{7,7}&P_{7,8}&P_{7,9}&P_{7,10}&P_{7,11}&P_{7,12}& & & & & & \\
 & & & & & & & & & & & &P_{8,1}&P_{8,2}&P_{8,3}&P_{8,4}&P_{8,5}&P_{8,6}&P_{8,7}&P_{8,8}&P_{8,9}&P_{8,10}&\textcolor{red}{P_{8,11}}&\textcolor{red}{P_{8,12}}& & & \\
 & & & & & & & & & & & & & & &P_{9,1}&P_{9,2}&P_{9,3}&P_{9,4}&P_{9,5}&P_{9,6}&P_{9,7}&\textcolor{red}{P_{9,8}}&\textcolor{red}{P_{9,9}}&\textcolor{red}{P_{9,10}}&\textcolor{red}{P_{9,11}}&\textcolor{red}{P_{9,12}}\\
 & & & & & & & & & & & & & & & & & &P_{10,1}&P_{10,2}&P_{10,3}&P_{10,4}&\textcolor{red}{P_{10,5}}&\textcolor{red}{P_{10,6}}&\textcolor{red}{P_{10,7}}&\textcolor{red}{P_{10,8}}&\textcolor{red}{P_{10,9}}\\
 & & & & & & & & & & & & & & & & & & & & &\textcolor{black}{P_{11,1}}&\textcolor{red}{P_{11,2}}&\textcolor{red}{P_{11,3}}&\textcolor{red}{P_{11,4}}&\textcolor{red}{P_{11,5}}&\textcolor{red}{P_{11,6}}\\
 & & & & & & & & & & & & & & & & & & & & & & & &\textcolor{red}{P_{12,1}}&\textcolor{red}{P_{12,2}}&\textcolor{red}{P_{12,3}}
 \end{array}
 \right]
\end{align*}
}
\caption{Parity-check matrix $H(\dl=3,\dr=6,L=9,M)$ of ($\dl,\dr,L$) codes. Each $P_{i,j}$ represents an $M\times M$ random permutation matrix. The red submatrix is used for termination. }
\label{h1}
\end{figure*}
\begin{figure*}
 \scriptsize{
 \begin{align*}
&\left[ \begin{array}{@{\hspace{0mm}}l@{\hspace{0mm}}l@{\hspace{0mm}}l@{\hspace{0mm}}l@{\hspace{0mm}}l@{\hspace{0mm}}l@{\hspace{0mm}}l@{\hspace{0mm}}l@{\hspace{0mm}}l@{\hspace{0mm}}l@{\hspace{0mm}}l@{\hspace{0mm}}l@{\hspace{0mm}}l@{\hspace{0mm}}l@{\hspace{0mm}}l@{\hspace{0mm}}l@{\hspace{0mm}}l@{\hspace{0mm}}l@{\hspace{0mm}}l@{\hspace{0mm}}l@{\hspace{0mm}}l@{\hspace{0mm}}l@{\hspace{0mm}}l@{\hspace{0mm}}l@{\hspace{0mm}}l@{\hspace{0mm}}l@{\hspace{0mm}}l@{\hspace{0mm}}l@{\hspace{0mm}}l@{\hspace{0mm}}l@{\hspace{0mm}}l@{\hspace{0mm}}l@{\hspace{0mm}}l@{\hspace{0mm}}l@{\hspace{0mm}}l@{\hspace{0mm}}l}
P_{1,10}&P_{1,11}&P_{1,12}& & & & & & & & & & & & & & & & & & & & & & & & \\
P_{2,7}&P_{2,8}&P_{2,9}&P_{2,10}&P_{2,11}&P_{2,12}& & & & & & & & & & & & & & & & & & & & & \\
P_{3,4}&P_{3,5}&P_{3,6}&P_{3,7}&P_{3,8}&P_{3,9}&P_{3,10}&P_{3,11}&P_{3,12}& & & & & & & & & & & & & & & & & & \\
P_{4,1}&P_{4,2}&P_{4,3}&P_{4,4}&P_{4,5}&P_{4,6}&P_{4,7}&P_{4,8}&P_{4,9}&P_{4,10}&P_{4,11}&P_{4,12}& & & & & & & & & & & & & & & \\
 & & &P_{5,1}&P_{5,2}&P_{5,3}&P_{5,4}&P_{5,5}&P_{5,6}&P_{5,7}&P_{5,8}&P_{5,9}&P_{5,10}&P_{5,11}&P_{5,12}& & & & & & & & & & & & \\
 & & & & & &P_{6,1}&P_{6,2}&P_{6,3}&P_{6,4}&P_{6,5}&P_{6,6}&P_{6,7}&P_{6,8}&P_{6,9}&P_{6,10}&P_{6,11}&P_{6,12}& & & & & & & & & \\
 & & & & & & & & &P_{7,1}&P_{7,2}&P_{7,3}&P_{7,4}&P_{7,5}&P_{7,6}&P_{7,7}&P_{7,8}&P_{7,9}&P_{7,10}&P_{7,11}&P_{7,12}& & & & & & \\
 & & & & & & & & & & & &P_{8,1}&P_{8,2}&P_{8,3}&P_{8,4}&P_{8,5}&P_{8,6}&P_{8,7}&P_{8,8}&P_{8,9}&P_{8,10}&\textcolor{black}{P_{8,11}}&\textcolor{black}{P_{8,12}}& & & \\
 & & & & & & & & & & & & & & &P_{9,1}&P_{9,2}&P_{9,3}&P_{9,4}&P_{9,5}&P_{9,6}&P_{9,7}&\textcolor{black}{P_{9,8}}&\textcolor{black}{P_{9,9}}&\textcolor{black}{P_{9,10}}&\textcolor{red}{P_{9,11}}&\textcolor{red}{P_{9,12}}\\
 & & & & & & & & & & & & & & & & & &P_{10,1}&P_{10,2}&P_{10,3}&P_{10,4}&\textcolor{black}{P_{10,5}}&\textcolor{black}{P_{10,6}}&\textcolor{black}{P_{10,7}}&\textcolor{red}{P_{10,8}}&\textcolor{red}{P_{10,9}}
 \end{array}
 \right].
 \end{align*}
 }
\caption{Parity-check matrix $\tilde{H}(\dl=3,\dr=6,L=9,M)$ of modified ($\dl,\dr,L$) codes.  The red submatrix is used for termination. }
\label{h2}
\end{figure*}
\section{Preliminaries}
In this section, we briefly review  $(\dl,\dr, L)$  codes introduced by Kudekar 
{\it et al.}~\cite{5695130}. We assume $\frac{\dr}{\dl}=:\dk\in\mathbb{Z}$ and $k\ge 2$.

The SC-LDPC codes are defined by the following protograph codes \cite{protograph}. 
The adjacency matrix of the protograph is referred to as a base matrix. 
The base matrix of ($\dl,\dr,L$)  code is given as follow. 
Let $H(\dl, \dr, L)$ be an $(L+\dl-1)\times \dk L$ band binary matrix 
of band size $\dr\times\dl$ and column weight $\dl$, where the band size is height $\times$ width of the band.
We refer to $L$ as coupling number. 
For example, $H(4,12,9)$ is given as
\begin{align*}
 \renewcommand\arraystretch{0.45}
 H(4,12,9)= \left[ \begin{array}{@{}l@{}l@{}l@{}l@{}l@{}l@{}l@{}l@{}l@{}l@{}l@{}l@{}l@{}l@{}l@{}l@{}l@{}l@{}l@{}l@{}l@{}l@{}l@{}l@{}l@{}l@{}l@{}l@{}l@{}l@{}l@{}l@{}l@{}l@{}l@{}l@{}l}
 1&1&1&&&&&&&&& \\
 1&1&1&1&1&1&&&&&&&& \\
 1&1&1&1&1&1&1&1&1&&&&&&& \\
 1&1&1&1&1&1&1&1&1&1&1&1&&&&& \\
 &&&1&1&1&1&1&1&1&1&1&1&1&1&&&& \\
 &&&&&&1&1&1&1&1&1&1&1&1&1&1&1&&& \\
 &&&&&&&&&1&1&1&1&1&1&1&1&1&1&1&1&& \\
 &&&&&&&&&&&&1&1&1&1&1&1&1&1&1&1&1&1& \\
 &&&&&&&&&&&&&&&1&1&1&1&1&1&1&1&1&1&1&1 \\
 &&&&&&&&&&&&&&&&&&1&1&1&1&1&1&1&1&1 \\
 &&&&&&&&&&&&&&&&&&&&&1&1&1&1&1&1 \\
 &&&&&&&&&&&&&&&&&&&&&&&&1&1&1 
 \end{array}
 \right].
\end{align*}

The protograph of the base matrix $H(4,12,9)$ is given in Fig.~\ref{bit1}. 
The protograph of $(\dl, \dr, L)$ codes have $\dk L$ variable nodes and $L+\dl-1$ check nodes.
The degree of variable nodes are all of degree $\dl=4$. On the other hand, the degree of check nodes are not uniform. 
check nodes at center are of degree $\dr=12$ and check nodes near boundaries have lower degree.

The Tanner graph of ($\dl,\dr,L$)  code is obtained by making $M$ copies of protographs of $H(\dl,\dr,L)$ and connecting edges among the same edge types. 
The parameter $M$ is referred to as lifting number. 
The parity check matrix $H(\dl,\dr,L,M)$ of a ($\dl,\dr,L$) code is given by replacing each 1 in $H(\dl,\dr,L)$ with an $M\times M$ random permutation matrix and 
each 0 with
an $M\times M$ zero matrix.
Let $P_{i,j}$ for $i \in [1,L+\dl-1]$, $j \in [1,\dr]$ be a binary random $M \times M$ permutation matrices. 
An example $H(4,12,9,M)$ is given in Fig.~\ref{h1}. 

Denote a code word of ($\dl,\dr,L$)  code by $(\underline{x}_1,\dotsc,\underline{x}_{kL})$, 
where $\underline{x}_{j}\in \mathbb{F}_2^{M}$ for $j\in [1,kL]$. 
Each $M$ rows of the $i$-th parity-check equations for $H(\dl,\dr,L,M)$ is written as
\begin{align}
&\sum_{j=1}^{\dr}P_{i,j}\underline{x}_{ik-\dr+j}=\underline{0} \label{131436_26Jun12}
\end{align}
for $i=1,\dotsc,L+\dl-1$. 
We assumed $\underline{x}_j =0$ for $j \notin [1,kL]$ for simplicity of notation. 

When the parity-check matrix $H$ of a protograph code is full-rank, the coding rate is given by 
$1-\frac{m}{n}$, with  $n$ variable nodes and $m$ check nodes.
Unless otherwise specified, we assume parity-check matrices are full-rank. 
Hence, the design coding rate of $(\dl,\dr,L)$  codes is given by
\begin{align}
 R&=\frac{\dk L-(L+\dl-1)}{\dk L}=\frac{\dk-1}{\dk}-\frac{\dl-1}{\dk L}.\label{084030_16Jul12}  
\end{align}
The rate-loss from the coding rate $(\dk-1)/\dk$ of $(\dl,\dr)$ codes is $\frac{\dl-1}{\dk L}$
which vanishes as $O(1/L).$ 

Lentmaier et al. \cite{5571910} observed that as $L$ increases, 
the BP threshold values $\epsilon^{\mathrm{BP}}$ of $(\dl,\dr,L)$ codes
approach the MAP threshold value of $(\dl,\dr)$ LDPC codes. 
Large coupling number $L$ is not preferred since the block error rate of SC-LDPC codes tends to 
be degraded when $L$ is large. 
Hence, mitigating the rate-loss while keeping the BP threshold is desired. 


\section{Issues in Encoding of Spatially-Coupled Codes}
\label{233249_20Jun12}
Encoding convolutional LDPC codes has been investigated by Pusane et al. in \cite{4568447}. 
Encoding of $(\dl,\dr,L)$ codes involves two parts, i.e., sequential encoding  and termination. 

For simplicity, we assume $M=1$ unless otherwise specified. 
Since there are $kL$ variable nodes and $L+\dl-1$ check nodes, 
the number of information bits $N_{\mathrm{info}}$ is given by 
\begin{align*}
N_{\mathrm{info}} = kL -(L+\dl-1). 
\end{align*}
In the sequential encoding stage, for each section $i=1,\dotsc,$ 
the parity bit $\underline{x}_{ik}$ is determined by 
the $i$-th check-node constraint \eqref{131436_26Jun12} and by using coded bits 
$\underline{x}_{1},\dotsc,\underline{x}_{(i-1)\dk}$ which was determined in the previous stages
and $\dk-1$ information bits $\underline{x}_{(i-1)k+1} , \dots , \underline{x}_{ik-1}$. 
The total number $N_{\mathrm{seq}}$ of parity bits which are determined in the sequential encoding stages 
is given by 
\begin{equation*}
N_\mathrm{seq} = \left\lfloor \frac{N_{\mathrm{info}}}{k-1} \right\rfloor=L-\left\lceil\frac{\dl-1}{k-1}\right\rceil. 
\end{equation*}
For the example of $H(4,12,9,M)$, $N_\mathrm{seq}(4,12)=7.$
We need to determine the remaining $N_\mathrm{term}$ undetermined parity bits 
from the remaining $N_\mathrm{info}-(k-1)N_\mathrm{seq}$ information bits, where
\begin{align*}
 N_{\mathrm{term}}=kL-(N_{\mathrm{info}}+N_\mathrm{seq})=\dl-1+\left\lceil\frac{\dl-1}{k-1}\right\rceil. 
\end{align*}
For the example of $H(4,12,9,M)$, $N_\mathrm{term}(4,12)=5.$

In Fig.~\ref{bit1}, protograph of $(4,12,9)$ code is shown. 
Gray nodes represent information bits. Light red nodes and red nodes are parity nodes determined 
in the sequential encoding stage and termination stage, respectively. 
\subsection{Sequential Encoding Stage}
Let us see how the $i$-th parity bit is determined in the sequential encoding stage. 
Define syndrome at section $i$ as follows. 
\begin{align}
\underline{s}_{i}&=\sum_{j=1}^{\dr-k}P_{i,j}\underline{x}_{ik-\dr+j},\label{120357_20Jun12}
\end{align}
where, for simplicity of notation, we define $\underline{x}_j=\underline{0}$ for $j\notin [1,kL]$. 
We start from $\underline{s}_1=\underline{0}$.
In sequential encoding part, for each section $i=1,2,\dotsc$, one parity bit node $\underline{x}_{k i}$ 
is sequentially 
determined from $k-1$ information bit nodes 
$\underline{x}_{k (i-1)+1},\dotsc, \underline{x}_{k i-1}$  and syndrome $\underline{s}_i$.
From \eqref{131436_26Jun12} and \eqref{120357_20Jun12}, we have 
\begin{align}
\underline{s}_{i}+\sum_{j=\dr-k+1}^{\dr}P_{i,j}\underline{x}_{ik-\dr+j}=\underline{0}. 
\end{align}
From this, $\underline{x}_{k i}$ can be determined as follows. 
\begin{align}
\underline{x}_{k i}= P_{i,\dr}^{-1}\bigl(\underline{s}_{i}+\sum_{j=\dr-k+1}^{\dr-1}P_{i,j}\underline{x}_{ik-\dr+j}\bigr).\label{130504_20Jun12} 
\end{align}
Since $P_{i,\dr}^{-1}$ is an $M\times M$ permutation matrix, \eqref{130504_20Jun12} can be accomplished with $O(M)$ computational costs. 
In Fig.~\ref{bit1}, protograph of $(4,12,9)$ code is shown. 
Sequential encoding is accomplished sequentially from left for each $\dk(=3)$ variable nodes.  
First, $\underline{x}_3$ is determined from $\underline{x}_{1},\underline{x}_{2}$. 
Next, $\underline{x}_6$ is determined from $\underline{s}_2,\underline{x}_4,\underline{x}_5$. 
Then $\underline{x}_9$ is determined from $\underline{s}_3,\underline{x}_7,\underline{x}_8$. 
This continues until $\underline{x}_{21}$ is determined from  $\underline{s}_{7},\underline{x}_{19},\underline{x}_{20}$. 
The sequential encoding process continues until the number of the remaining information bit nodes 
reaches less than $k-1$. 
After that, we move to termination stage. 
\subsection{Termination}
In the termination stage, we determine the remaining $N_{\mathrm{term}}$ parity bits 
so that parity-check equations \eqref{131436_26Jun12} are satisfied for 
$i\in [N_\mathrm{seq}+1,N_{\mathrm{seq}}+N_\mathrm{term}]$. 
We define 
the $i$-th syndrome $\underline{s}_i$ ($i\in [N_\mathrm{seq}+1,N_\mathrm{seq}+N_\mathrm{term}]$) by
 the contribution from determined bits to the $i$-the check nodes. 
To be precise, 
\begin{align*}
 \underline{s}_i :&= \sum_{j':=ik-\dr+j\le N_\mathrm{info}+N_\mathrm{seq}} P_{i,j}\underline{x}_{j'}\\
 & = \sum_{j=1}^{\dr-i\dk+N_\mathrm{info}+N_\mathrm{seq}} P_{i,j}\underline{x}_{ik-\dr+j}. 
\end{align*}
Denote  the right bottom $N_\mathrm{term}M\times N_\mathrm{term}M$ submatrix of $H(\dl,\dr,L,M)$ by $H_\mathrm{term}$.
The termination process of $(\dl,\dr,L)$ codes is equivalent to solve the following equation
\begin{align}
&H_\mathrm{term} (\underline{x}_{kL-N_\mathrm{term}+1},\dotsc,\underline{x}_{kL})^T\nonumber\\
&\quad = (\underline{s}_{L+\dl-N_\mathrm{term}},\dotsc,\underline{s}_{L+\dl-1})^T\label{214226_28Jun12}. 
\end{align}
When $H_\mathrm{term}$ is not full-rank, one can modify some entries of $H_\mathrm{term}$ so that $H_\mathrm{term}$ becomes full-rank 
and the decoding performance remains  almost the same.
The inverse of $H_\mathrm{term}$ is not sparse in general. 
Hence, one needs $O(M^2)$ computational cost to solve the linear equations \eqref{214226_28Jun12}.

\begin{figure*}[t]
\setlength\unitlength{1truecm}
\begin{picture}(17,3.5)(0,0)
 \put(0.0,0.5){\includegraphics[scale=0.7]{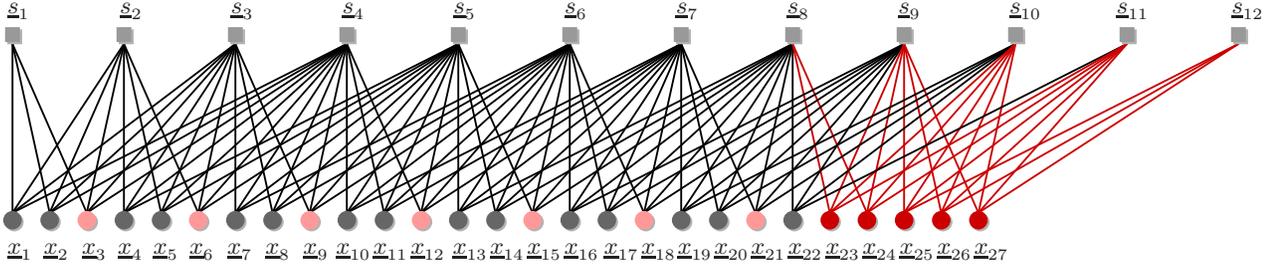}}
\put(0.18,0.3){{\small$\underline{x}_1 \hspace{0.155cm}\underline{x}_2 \hspace{0.175cm}\underline{x}_3 \hspace{0.145cm}\underline{x}_4 \hspace{0.145cm}\underline{x}_5 \hspace{0.145cm}\underline{x}_6 \hspace{0.175cm}\underline{x}_7 \hspace{0.175cm}\underline{x}_8 \hspace{0.175cm}\underline{x}_9 \hspace{0.105cm}\underline{x}_{10} \hspace{0.035cm}\underline{x}_{11} \hspace{0.035cm}\underline{x}_{12} \hspace{0.1cm}\underline{x}_{13} \hspace{0.035cm}\underline{x}_{14} \hspace{0.035cm}\underline{x}_{15} \hspace{0.035cm}\underline{x}_{16} \hspace{0.07cm}\underline{x}_{17} \hspace{0.035cm}\underline{x}_{18} \hspace{0.025cm}\underline{x}_{19} \hspace{0.025cm}\underline{x}_{20} \hspace{0.03cm}\underline{x}_{21} \hspace{0.035cm}\underline{x}_{22} \hspace{0.035cm}\underline{x}_{23} \hspace{0.035cm}\underline{x}_{24} \hspace{0.035cm}\underline{x}_{25} \hspace{0.035cm}\underline{x}_{26} \hspace{0.035cm}\underline{x}_{27}$}}
\put(0.18,3.5){{\small$\underline{s}_1 \hspace{1.2cm}\underline{s}_2\hspace{1.18cm}\underline{s}_3\hspace{1.18cm}\underline{s}_4\hspace{1.18cm}\underline{s}_5\hspace{1.18cm}\underline{s}_6\hspace{1.18cm}\underline{s}_7\hspace{1.18cm}\underline{s}_8\hspace{1.18cm}\underline{s}_9\hspace{1.18cm}\underline{s}_{10}\hspace{1.0cm}\underline{s}_{11}\hspace{1.1cm}\underline{s}_{12}$ }}
\end{picture}
\caption{Protograph of $(\dl=4,\dr=12,L=9)$ codes. Gray nodes represent information bits. Light red nodes represent parity bits. Red nodes represent parity bits calculated in termination. 
Sequential encoding is accomplished sequentially from left for each $\dk(=3)$ nodes. 
Termination requires to solve linear equations involving as much as five variable nodes $\underline{x}_{23},\dotsc, \underline{x}_{27}$ shown in \eqref{124506_20Jun12}.
One needs $O(M^2)$ computational cost to solve the linear equations. 
}
\label{bit1}
\end{figure*}

For the example of $H(4,12,9,M)$, we need to solve linear equations 
$H_\mathrm{term}(\underline{x}_{23},\dotsc,\underline{x}_{27})^T=(\underline{s}_{8},\dotsc,\underline{s}_{12})^T$ with
{\small
\begin{align}
H_\mathrm{term}=\left(\begin{array}{ccccc}
 P_{8,11}&P_{8,12}&&&\\
P_{9,8}&P_{9,9}&P_{9,10}&P_{9,11}&P_{9,12}\\
P_{10,5}&P_{10,6}&P_{10,7}&P_{10,8}&P_{10,9}\\
P_{11,2}&P_{11,3}&P_{11,4}&P_{12,5}&P_{11,6}\\
&&P_{12,1}&P_{12,2}&P_{12,3}
\end{array} 
\right)
\label{124506_20Jun12}
\end{align}
}
In Fig.~\ref{h1}, $H_\mathrm{term}$ is written in red in $H(4,12,9,M)$. 
In Fig.~\ref{bit1} red nodes represent bit node involved in the termination process. 
\begin{figure*}[t]
\setlength\unitlength{1truecm}
\begin{picture}(17,3.5)(0,0)
 \put(0.0,0.5){\includegraphics[scale=0.7]{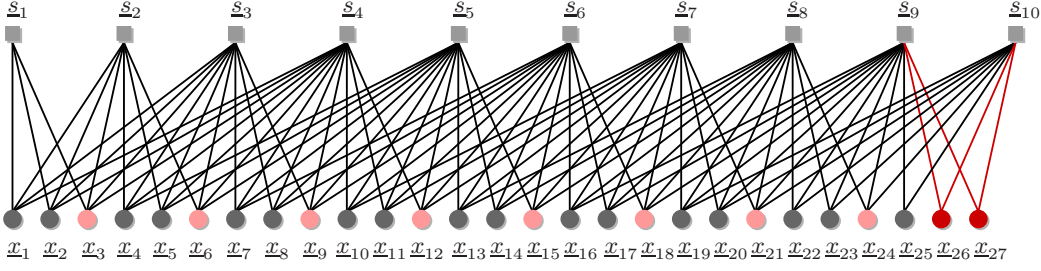}}
\put(0.18,0.3){{\small$\underline{x}_1 \hspace{0.155cm}\underline{x}_2 \hspace{0.175cm}\underline{x}_3 \hspace{0.145cm}\underline{x}_4 \hspace{0.145cm}\underline{x}_5 \hspace{0.145cm}\underline{x}_6 \hspace{0.175cm}\underline{x}_7 \hspace{0.175cm}\underline{x}_8 \hspace{0.175cm}\underline{x}_9 \hspace{0.105cm}\underline{x}_{10} \hspace{0.035cm}\underline{x}_{11} \hspace{0.035cm}\underline{x}_{12} \hspace{0.1cm}\underline{x}_{13} \hspace{0.035cm}\underline{x}_{14} \hspace{0.035cm}\underline{x}_{15} \hspace{0.035cm}\underline{x}_{16} \hspace{0.07cm}\underline{x}_{17} \hspace{0.035cm}\underline{x}_{18} \hspace{0.025cm}\underline{x}_{19} \hspace{0.025cm}\underline{x}_{20} \hspace{0.03cm}\underline{x}_{21} \hspace{0.035cm}\underline{x}_{22} \hspace{0.035cm}\underline{x}_{23} \hspace{0.035cm}\underline{x}_{24} \hspace{0.035cm}\underline{x}_{25} \hspace{0.035cm}\underline{x}_{26} \hspace{0.035cm}\underline{x}_{27}$}}
\put(0.18,3.5){{\small$\underline{s}_1 \hspace{1.2cm}\underline{s}_2\hspace{1.18cm}\underline{s}_3\hspace{1.18cm}\underline{s}_4\hspace{1.18cm}\underline{s}_5\hspace{1.18cm}\underline{s}_6\hspace{1.18cm}\underline{s}_7\hspace{1.18cm}\underline{s}_8\hspace{1.18cm}\underline{\tilde{s}}_9\hspace{1.18cm}\underline{\tilde{s}}_{10}$ }}
\end{picture}
\caption{Protograph of modified $(\dl=4,\dr=12,L=9)$ codes. This protograph is produced by removing ($\dl-2$) right most check nodes from the protograph of $(\dl=4,\dr=12,L=9)$ codes. 
Termination is accomplished by solving linear equations involving only two variable nodes $\underline{x}_{26}$ and $\underline{x}_{27}$.
One can solve the linear equations by an accumulator with $O(M)$ computational costs like RA codes. 
}
\label{bit2}
\end{figure*}
\section{Modified Spatially-Coupled Codes}
In this section, we propose modified ($\dl,\dr,L$) codes. We give an explanation how the modified codes can be  efficiently terminated. 
\subsection{Modified ($\dl,\dr,L$) Codes}
The base matrix $\tilde{H}(\dl,\dr,L)$ of modified $(\dl,\dr,L)$ codes is obtained by removing $\dl-2$ bottom rows of the base matrix ${H}(\dl,\dr,L)$ of 
$(\dl,\dr,L)$ codes. 
For example, $\tilde{H}(4,12,9)$ is given as follows. 
\begin{align*}
 \renewcommand\arraystretch{0.45}
 \tilde{H}(4,12,9)= \left[ \begin{array}{@{}l@{}l@{}l@{}l@{}l@{}l@{}l@{}l@{}l@{}l@{}l@{}l@{}l@{}l@{}l@{}l@{}l@{}l@{}l@{}l@{}l@{}l@{}l@{}l@{}l@{}l@{}l@{}l@{}l@{}l@{}l@{}l@{}l@{}l@{}l@{}l@{}l}
 1&1&1&&&&&&&&& \\
 1&1&1&1&1&1&&&&&&&& \\
 1&1&1&1&1&1&1&1&1&&&&&&& \\
 1&1&1&1&1&1&1&1&1&1&1&1&&&&& \\
 &&&1&1&1&1&1&1&1&1&1&1&1&1&&&& \\
 &&&&&&1&1&1&1&1&1&1&1&1&1&1&1&&& \\
 &&&&&&&&&1&1&1&1&1&1&1&1&1&1&1&1&& \\
 &&&&&&&&&&&&1&1&1&1&1&1&1&1&1&1&1&1& \\
 &&&&&&&&&&&&&&&1&1&1&1&1&1&1&1&1&1&1&1 \\
 &&&&&&&&&&&&&&&&&&1&1&1&1&1&1&1&1&1 
 \end{array}
 \right]
\end{align*}

The protograph of modified codes is obtained by removing ($\dl-2$) right most check nodes and their incident edges from the protograph of $(\dl,\dr,L)$ codes.
In Fig.~\ref{bit2}, for example, we show protograph of modified $(\dl=4,\dr=12,L=9)$ codes. 
The coding rate of the modified $(\dl,\dr,L)$ codes is give by
\begin{equation}
\tilde{R}(\dl,\dr,L) = \frac{kL-(L+1)}{kL} = \frac{k-1}{k} -\frac{1}{kL}\label{215732_28Jun12}. 
\end{equation}
The rate-loss from the rate $\frac{\dk-1}{\dk}$ of $(\dl,\dr)$ codes is $\frac{1}{kL}$. 
Comparing \eqref{084030_16Jul12} and \eqref{215732_28Jun12}, 
we can see that rate-loss is mitigated as $(\dl-1)$-times as much by modifying. 

\begin{table*}[t]
 \caption{
Comparison of coding rate and BP threshold values for modified $(\dl,\dr,\dL)$ codes and $(\dl,\dr,\dL)$ codes.
}
\begin{center}
  \begin{tabular}{c|cccc} 
 $(\dl,\dr,\dL)$ & $\tilde{\epsilon}^{\mathrm{BP}}$ & $\epsilon^{\mathrm{BP}}$ & $\tilde{R}$ & $R$  \\\hline
 $(3,6,9)$ & 0.49174 & 0.51203 & 0.44444 & 0.38889  \\
 $(3,6,17)$ & 0.48816 & 0.48876 & 0.47059 & 0.41177  \\
 $(3,6,33)$ & 0.48815 & 0.48815 & 0.48485 & 0.46970  \\
 $(3,6,65)$ & 0.48815 & 0.48815 & 0.49231 & 0.48462  \\ \hline
 $(4,8,9)$ & 0.50158 & 0.51938 & 0.44444 & 0.33333  \\
 $(4,8,17)$ & 0.49774 & 0.49787 & 0.47059 & 0.41177  \\
 $(4,8,33)$ & 0.49774 & 0.49774 & 0.48485 & 0.45455  \\
 $(4,8,65)$ & 0.49774 & 0.49774 & 0.49231 & 0.47692  
 \end{tabular}
 \begin{tabular}{c|cccc} 
 $(\dl,\dr,\dL)$ & $\tilde{\epsilon}^{\mathrm{BP}}$ & $\epsilon^{\mathrm{BP}}$ & $\tilde{R}$ & $R$  \\\hline
 $(3,9,9)$ & 0.32157 & 0.33305& 0.62963 & 0.59259  \\
 $(3,9,17)$ & 0.31997 & 0.31995 & 0.64706 & 0.62745  \\
 $(3,9,33)$ & 0.31965 & 0.31965 & 0.65657 & 0.64647  \\
 $(3,9,65)$ & 0.31965 & 0.31965 & 0.66154 & 0.65641  \\ \hline
 $(4,12,9)$ & 0.33282 & 0.33282 & 0.62963 & 0.52963  \\
 $(4,12,17)$ & 0.33025 & 0.33033 & 0.64706 & 0.60784  \\
 $(4,12,33)$ & 0.33025 & 0.33025 & 0.65657 & 0.63636  \\
 $(4,12,65)$ & 0.33025 & 0.33025 & 0.66154 & 0.65128  \\
 \end{tabular}
\end{center}
\label{bpr}
\end{table*}
\section{Efficient Encoding of Modified Codes}
In this section, we explain how modified ($\dl,\dr,L$) codes can be efficiently encoded. 
\subsection{Sequential Encoding}
Sequential encoding is performed in the same way as $(\dl,\dr,L)$ codes. 
The number $\tilde{N}_{\mathrm{info}}$ of information bits of modified $(\dl,\dr,L)$ codes is given by
\begin{align*}
\tilde{N}_{\mathrm{info}} = kL -(L+1). 
\end{align*}
The total number $\tilde{N}_{\mathrm{seq}}$ of parity-bits which are determined in the sequential encoding stages 
is given by 
\begin{equation*}
\tilde{N}_\mathrm{seq} = \left\lfloor\frac{\tilde{N}_\mathrm{info}}{\dk-1}\right\rfloor=L-1
\end{equation*}
\subsection{Termination}
It follows that the number of parity bit nodes we need to determine in termination process is 
\begin{equation*}
\tilde{N}_{\mathrm{term}} = kL - \tilde{N}_{\mathrm{info}} - \tilde{N}_{\mathrm{seq}} = 2.
\end{equation*}
Note that $\tilde{N}_{\mathrm{term}}(=2)$ does not depend on $\dl$ and $\dr$.
This means that we always need deal with only two parity bits to terminate the encodin process. 
In the termination stage, we determine the remaining $\tilde{N}_{\mathrm{term}}(=2)$ parity bits 
so that parity-check equations \eqref{131436_26Jun12} are satisfied for 
$i\in \{\tilde{N}_\mathrm{seq}+1,\tilde{N}_\mathrm{seq}+2\}=\{L,L+1\}$. 

We define the $i$-th syndrome $\underline{\tilde{s}}_i\in \mathbb{F}_2^M\ (i\in \{L,L+1\})$ by
 the contribution from determined  bits to the $i$-the check nodes. 
To be precise, 
\begin{align*}
 \underline{\tilde{s}}_i  & = \sum_{j=1}^{\dr-i\dk+\tilde{N}_\mathrm{info}+\tilde{N}_\mathrm{seq}} P_{i,j}\underline{x}_{ik-\dr+j}\\
  & = \sum_{j=1}^{\dr-i\dk+kL-2} P_{i,j}\underline{x}_{ik-\dr+j}.
\end{align*}
Denote  the right bottom $\tilde{N}_\mathrm{term}M\times \tilde{N}_\mathrm{term}M$ submatrix of $\tilde{H}(\dl,\dr,L,M)$ by 
$\tilde{H}_\mathrm{term}$.
The termination process of modified $(\dl,\dr,L)$ codes is equivalent to solve the following linear equation.
\begin{align}
&\tilde{H}_\mathrm{term} (\underline{x}_{kL-1},\underline{x}_{kL})^T
 = (\underline{\tilde{s}}_{L},\underline{\tilde{s}}_{L+1})^T\label{222326_28Jun12},\\
&\tilde{H}_\mathrm{term} = 
 \left[ \begin{array}{ll}
  P_{L,\dl\dk-1}&P_{L,\dl\dk} \\
		 P_{L+1,(\dl-1)\dk-1}&P_{L+1,(\dl-1)\dk}
		\end{array}
 \right].\notag
\end{align}
In Fig.~\ref{h2}, the submatrix $\tilde{H}_\mathrm{term}$ is printed in red. 
In Fig.~\ref{bit2},  parity bits involved in termination are drawn in red. 

\begin{figure*}[t]
\setlength\unitlength{1truecm}
\begin{picture}(17,5.5)(0,0)
 \put(0.3, 3){\rotatebox{90}{$p_{i}^{(\ell)}$}}
\put(0.5,0){  \includegraphics[width=0.45\textwidth]{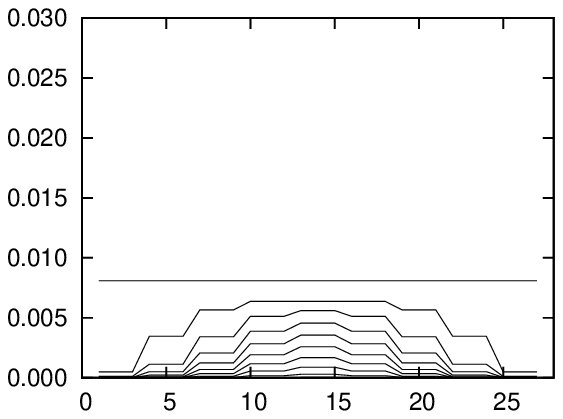} }
 \put(8.8, 3){\rotatebox{90}{$\tilde{p}_{i}^{(\ell)}$}}
\put(9,0){ \includegraphics[width=0.45\textwidth]{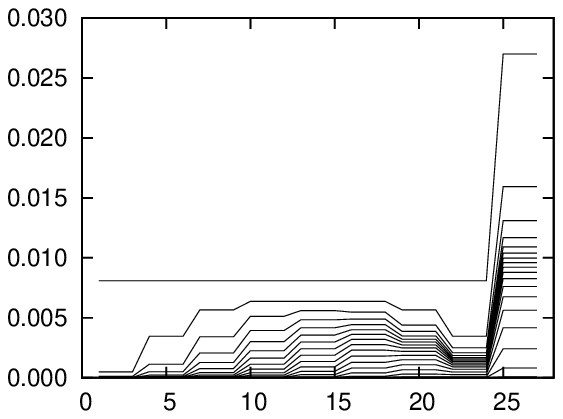} }
\end{picture} 
\caption{Comparison of convergence speed of decoding erasure rate ${p}_{i}^{(\ell)}$ and $\tilde{p}_{i}^{(\ell)}$ at the $\ell$-th BP iteration at section $i\in [1,27]$ for (4,12,9) codes (left) and modified (4,12,9) codes (right)
BEC($\epsilon$=0.3)}
 \label{224137_28Jun12}
\end{figure*}

Since the sum of all rows of $\tilde{H}_\mathrm{term}$ is $\underline{0}\in \mathbb{F}_2^{2M}$, 
$\tilde{H}_\mathrm{term}$ is not full-rank. 
We need a slight modification to ensure that  
$\underline{x}_{kL-1}$ and $\underline{x}_{kL}$ can be determined from $\underline{\tilde{s}}_{L}$ and $\underline{\tilde{s}}_{L+1}$.
We modify $\tilde{H}_\mathrm{term}$ as follows.
\begin{align}
&\tilde{H}_\mathrm{term}:= 
\left[ \begin{array}{ll}
I_M&I_M' \\
I_M&I_M
 \end{array}
 \right],\label{165809_16Jul12}\\
& I_M'=
\left[ \begin{array}{@{\hspace{0mm}}l@{\hspace{3mm}}l@{\hspace{3mm}}l@{\hspace{3mm}}l@{\hspace{3mm}}l@{\hspace{0mm}}l}
0&0&0&0&0\\
1&0&0&0&0\\
0&1&0&{\ddots}&0\\
0&0&1&0&0\\
0&0&0&1&0
 \end{array}
 \right]\in \mathbb{F}_2^{M\times M},\notag
\end{align}
where $I_{M}$ is the identity matrix of size $M \times M$. 
The effect for decoding performance from this modification can be negligibly small when lifting number $M$ is large. 
The equation \eqref{222326_28Jun12} with $\tilde{H}_\mathrm{term}$ \eqref{165809_16Jul12}
can be solved by $2M$ accumulations with an accumulator in the following way. 
Let us define
$ \underline{x}_{kL-1}=:(x_{kL-1,1},\dotsc,x_{kL-1,M})$,
$ \underline{x}_{kL}=:(x_{kL,1},\dotsc,x_{kL-1,M}). $
The solution $\underline{x}_{kL-1}$ and $\underline{x}_{kL1}$ in 
the equation \eqref{222326_28Jun12} can be solved by 
sequentially calculating
\begin{align*}
 x_{kL-1,i}&=x_{kL,i-1}  +s_{L,i}\\
 x_{kL,i}  &=x_{kL-1,i}+s_{L+1,i} 
\end{align*}
for $i=1\dotsc,M$, where we assumed $x_{kL,-1}=0$. 
Thus, the modified $(\dl,\dr,L)$ codes can be terminated with computational costs $O(M)$.

\section{BP Threshold}
In this section, we give comparison of BP threshold values for $(\dl,\dr,L)$ codes and modified $(\dl,\dr,L)$ codes over the binary erasure channel 
(BEC) with erasure probability $\epsilon$. 

Table \ref{bpr} shows BP threshold values $\epsilon^{\mathrm{BP}}(\dl,\dr,L)$ (resp.~$\tilde{\epsilon}^{\mathrm{BP}}(\dl,\dr,L)$)
coding rate $R(\dl,\dr,L)$ (resp.~$R(\dl,\dr,L)$) of $(\dl,\dr,L)$ codes (resp.~ modified $(\dl,\dr,L)$ codes). 
$\epsilon^\mathrm{Sha}$ represents the Shannon threshold for coding rate $1-\dl/\dr$. 
It can be seen that modified $(\dl,\dr,L)$ codes have  higher coding rate and the same threshold as $(\dl,\dr,L)$ codes within the 5 digits precision. 

Let $p_i^{(\ell)}$ and $\tilde{p}_i^{(\ell)}$ be the decoding erasure rate of 
$\underline{x}_i$ at the $\ell$-th BP iteration for $(4,12,9)$ codes and modified $(4, 12,9)$ codes, respectively. 
Figure \ref{224137_28Jun12} shows how $p_i^{(\ell)}$ and $\tilde{p}_i^{(\ell)}$ converge to $\underline{0}$. 
It can be seen that modified codes have slower convergence.
\section{Conclusion}
We propose modified spatially-coupled codes.
The modified codes have less rate-loss,  and have the same threshold as $(\dl,\dr,L)$ codes and are efficiently terminable by using an accumulator. 

Future works include analysis of convergense, finite length performance, and performance over other channels like \cite{KuKaMAC} and \cite{KuKaDEC}. 
\bibliographystyle{IEEEtran} 
\bibliography{IEEEabrv,../../kenta_bib}
\end{document}